\documentclass[11pt]{article}
\usepackage{amssymb}
\usepackage{amsfonts}
\usepackage{amsmath}
\usepackage{graphicx}
\usepackage{color}
\begin{document}

\title{Are human interactivity times lognormal?}
\author{Norbert Blenn\thanks{Faculty of Electrical Engineering, Mathematics and
Computer Science, P.O Box 5031, 2600 GA Delft, The Netherlands; \emph{email}:
N.Blenn@, P.F.A.VanMieghem@tudelft.nl} and Piet Van Mieghem}
\date{Delft University of Technology\\
16 December 2015}
\maketitle

\begin{abstract}
In this paper, we are analyzing the interactivity time, defined as the duration between two consecutive tasks such as sending emails, collecting friends and followers and writing comments in online social networks (OSNs). The distributions of these times are heavy tailed and often described by a power-law distribution.
However, power-law distributions usually only fit the heavy tail of empirical data and ignore the information in the smaller value range. Here, we argue that the durations between writing emails or comments, adding friends and receiving followers are likely to follow a lognormal distribution.

We discuss the similarities between power-law and lognormal distributions, show that binning of data can deform a lognormal to a power-law distribution and propose an explanation for the appearance of lognormal interactivity times. The historical debate of similarities between lognormal and power-law distributions is reviewed by illustrating the resemblance of measurements in this paper with the historical problem of income and city size distributions.
\end{abstract}

\section{Introduction}

\label{sec_introduction}

Massive data from online social media and online social networking (OSN) enables accurate analysis of the human behavior and of the interaction times between individuals through technology such as in word-of-mouth marketing, opportunistic networking and viral spreading. Empirical measurements contradict the commonly assumed exponential distribution of Poissonian inter-event times \cite{PVM_PAComplexNetsCUP} in spreading processes or epidemic models. Multiple publications \cite{Barabasi05nature,Eckmann2004,goncalves2008,Radicchi2008a,Leskovec_msn2008,Candia2008,Karsai_NSR2012} report that \emph{the interactivity time}, defined as the duration between two consecutive tasks like sending emails, accessing web pages, instant messaging and phone calls, follow power-law distributions. These findings recently led to non-Markovian analyses, addressed in the work of Cator \textit{et al.} \cite{Cator2013}, Iribarren and Moro \cite{Iribarren09}, Van Mieghem and van de Bovenkamp \cite{VanMieghem2013} and Schweizer \textit{et al.} \cite{Schweitzer2014}. Given that the inter-activity distributions are heavy tailed, word-of-mouth spreading, viral infections or dynamics of memes are expected to endure or survive longer, compared to basic Markovian models \cite{Iribarren09,Karsai2011}. Barabasi \cite{Barabasi05nature} infers that heavy-tailed distributions may arise from a priority queue, where individuals execute tasks of which the majority can be completed in short time, but some tasks wait long due to a perceived priority. Barabasi's priority queue model fits the distribution of durations between events quite well, leading to a power-law distribution with an exponent $\gamma$ around 1.

A power-law random variable $X\geq\tau$ has the probability density function
\begin{equation}
f_{X}(t)=ct^{-\gamma}\hspace{2cm}t\geq\tau
\label{pdf_powerlaw}
\end{equation}
where $c=\frac{1-\gamma}{\tau^{1-\gamma}}$ and $\tau>0$ is the lower bound for $X$. The probability density function (pdf) of a lognormal random variable $X$ for $t\geq0$ is 
\begin{equation}
f_{X}\left(  t\right)  =\frac{\exp\left[  -\frac{(\log t-\mu)^{2}}{2\sigma^{2}}\right]  }{\sigma t\sqrt{2\pi}} 
\label{pdf_lognormal}
\end{equation}
where $\left(  \mu,\sigma\right)  $ are called the parameters of the lognormal pdf, that are the mean and variance of $\log X$ as shown in Appendix \ref{sec_properties_lognormal}.

When we assume that inter-event durations are power-law distributed, we encounter the following issues:
\begin{enumerate}
\item In many cases, only a part of the data (the tail larger than $\tau$) is modeled by a power-law. The lower bound $\tau$ in (\ref{pdf_powerlaw}) does not correspond to the physical minimum of the random variable $X$, but $\tau$ is fitted from the data by ignoring smaller values that do not obey the power-law. Often, these smaller values may have a large probability to occur, so that their neglect is difficult to justify. In other words, only a part of the process (above $\tau$) is modeled by a power-law (\ref{pdf_powerlaw}), while the other part (below $\tau$) is not.

\item Most processes or measurements possess both a lower as well as an upper bound. Apart from the lower bound $\tau$, an upper bound $\kappa$ is often invoked, at which a cut-off is observed: the power-law behavior is confined to the range $\tau\leq X\leq\kappa$, although $X_{\min}<\tau$ and $X_{\max }>\kappa$. However, it is often unclear whether the process in the deep tail still obeys a power-law distribution or some other, much faster decreasing distribution. The upper bound $\kappa$ is usually empirically determined, rather than based on the physical maximum of $X$. As long as \[\Pr[X>\kappa]=c\int_{\kappa}^{\infty}t^{-\gamma}dt=\left(  \frac{\kappa}{\tau}\right)  ^{1-\gamma}\] is small (with respect to the measurement precision), the upper bound $\kappa$ is justified, else other validation arguments are needed.

\item We demonstrate here that the binning of data (either by the data-provider or by the researcher) alters the shape of a lognormal distribution into an apparent power-law.
\end{enumerate}

Particularly in relation to human activities or behaviors, we question in this paper the widely assumed power-law distribution. 

We present measurements of inter-event durations from Digg.com and Reddit.com, two online social news aggregators \cite{Tang2011,Doerr-etal:2012,VanMieghem11}, and from the Enron data set\footnote{Enron Email Data-set, Leslie Kaelbling and Melinda Gervasio, http://www.cs.cmu.edu/\~{ }enron/}, a collection of emails sent by employees of the company Enron, and argue that a lognormal distribution is a valid candidate for the distribution of human inter-event durations in Section \ref{Sec_measurements}. The problem of fitting a lognormal is explained in Section \ref{Sec_fittinglognormal}, followed by previously reported lognormally distributed data sets in Section \ref{Sec_reported_parameters}. Existing theoretical models are compared in Section \ref{Sec_debate}, a plausible interpretation for lognormal human behavior is proposed in Section \ref{sec_plausible_explanation_lognormal_interactivity_times} and Section \ref{sec_conclusion} concludes. Mathematical properties of the lognormal distribution are deferred to the Appendix \ref{sec_properties_lognormal}. 
Appendix \ref{sec_likelihood_testing} presents results of likelihood tests for the observed distributions.

\section{Observations and Measurements from OSN}

\label{Sec_measurements}

All events in an OSN are based on users' activities, such as posts, friend-requests and comments. A complete data set including all activities of users from Digg.com for a duration of 4 years, described in Tang \textit{et al.} \cite{Tang2011} and Doerr \textit{et al.} \cite{Doerr-etal:2012}, allows us to analyze the time frame in which users of Digg.com add their friends. Doerr \textit{et al.} \cite{doerr2013-infection} found that reaction times in a retweet network from Twitter and Digg are close to a lognormal distribution with parameters $\mu=10.1$ and $\sigma=2.2$. 
The process of adding friends shows bursts of activity also observed and analyzed in email communication \cite{Barabasi05nature,Iribarren09, Stouffer:2005vn, Malmgren2008, Malmgren2009}. In these publications the question arises, whether the observed distribution of inter-activity times is described by a power-law, lognormal or a cascading Poisson process.

Malmgren \textit{et al.} \cite{Malmgren2008,Malmgren2009} describe that circadian and weekly activity cycles in human behavior are the factors that lead to heavy tails. They proposed a cascading Poisson process, consisting of a {\em nonhomogeneous} Poisson process that reflects periodicity and a {\em homogeneous} Poisson process describing active intervals, which is shown to model the interactivity distributions of e-mail communication. 

Because the network of Digg.com is directed (like in Twitter.com or other OSNs), a user can be followed by other users to become their \textquotedblleft friend\textquotedblright, while a user cannot add followers. This means that the process of adding friends is solely based on the user himself, whereas obtaining followers depends on the activities of other users. The random variable $T_{friend}$ denotes the time between the addition of two friends and, similarly, $T_{follower}$ is the time between receiving two followers.%
\begin{figure}[ht]
\begin{center}
\includegraphics[width=1.00\linewidth]{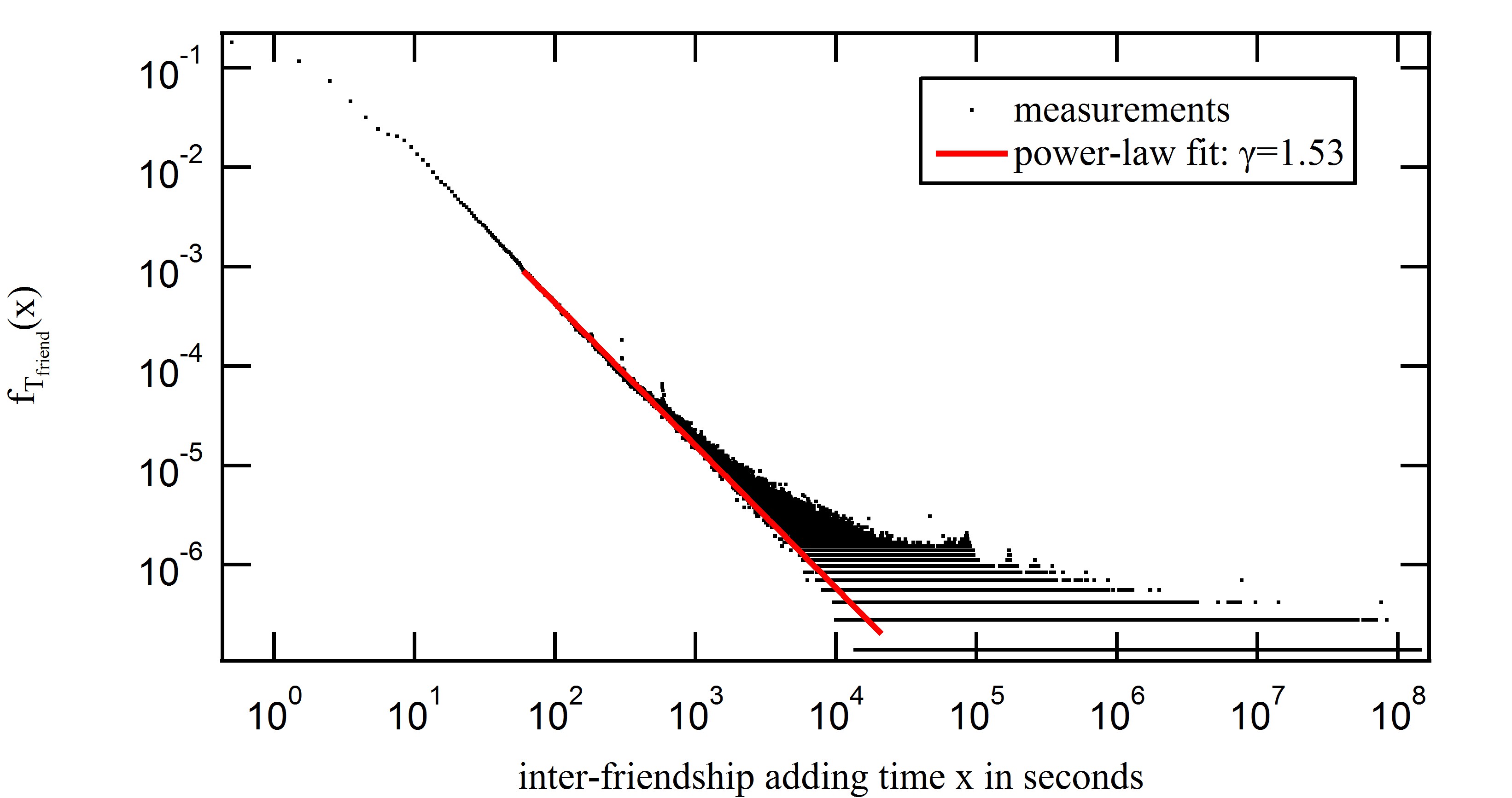}%
\caption{Time difference (in seconds) between the addition of two consecutive friends.}
\label{Fig_interfriendaddtime_seconds}
\end{center}
\end{figure}

\begin{figure}[ht]
\begin{center}
\includegraphics[width=1.00\linewidth]{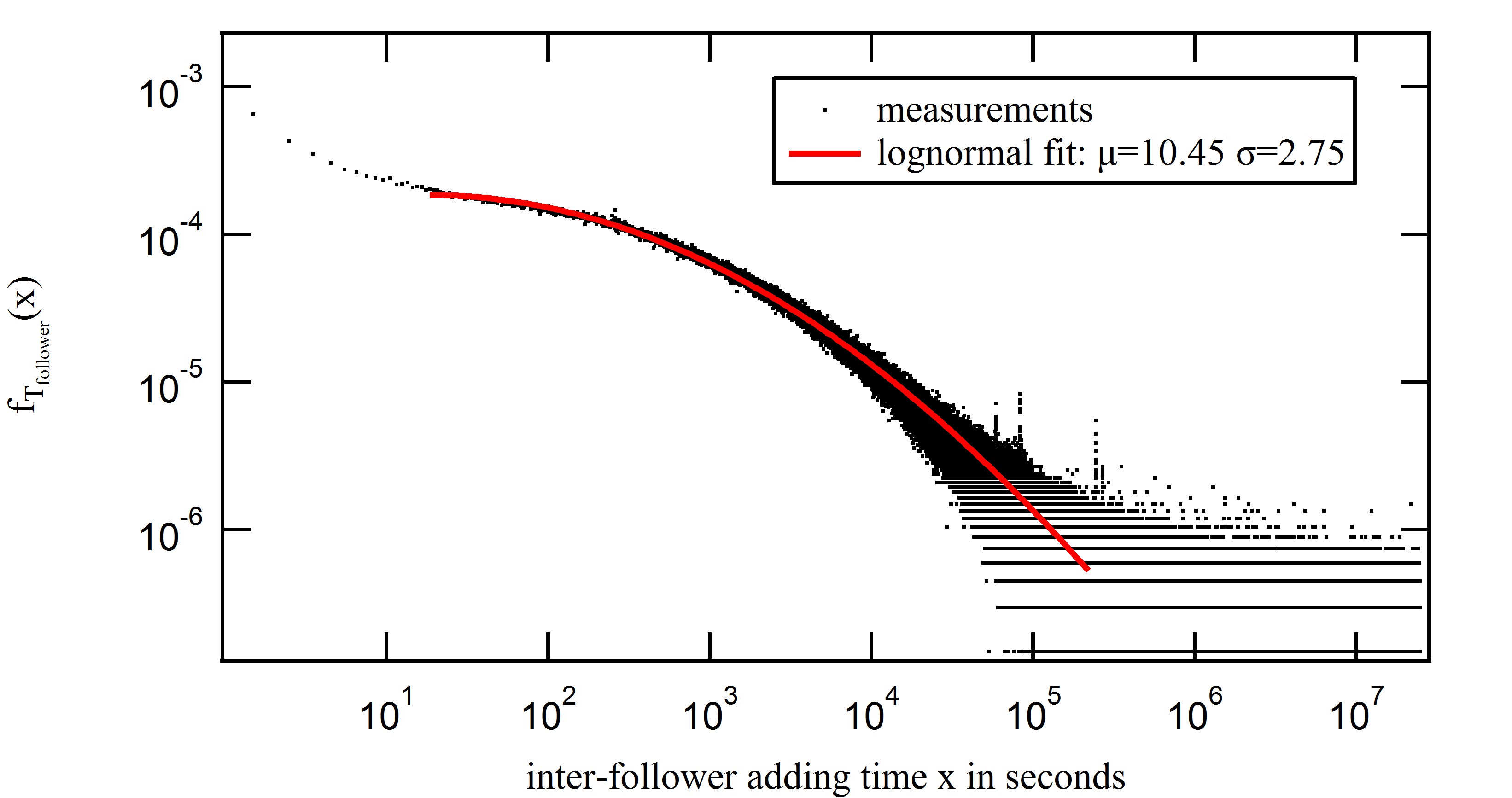}
\caption{Time differences (in seconds) between the receipt of two consecutive followers.}
\label{Fig_binsize1s_followers}
\end{center}
\end{figure}

Figures \ref{Fig_interfriendaddtime_seconds} and \ref{Fig_binsize1s_followers} depict the distribution of durations between adding friends $T_{friend}$ and of receiving followers $T_{follower}$ for 7.4 million friendship relations in Digg.  Fitting the data, i.e. all realizations of $T_{friend}$, by the state of the art technique by Clauset \textit{et al.} \cite{Clauset:2009} to a power-law (\ref{pdf_powerlaw}) results in an exponent $\gamma=1.53$ for $\tau=59s$, whereas all realizations of $T_{follower}$ are fitted best by a lognormal (\ref{pdf_lognormal}) with parameters $\mu=10.45$ and $\sigma=2.75$. For $T_{follower}$ an estimated p-value of 0.0 indicates that a power-law fit provides not the best solution, whereas the p-value for $T_{friend}$ of $0.23$ indicates a reasonable fit for a power-law distribution.%

The reason that the distribution of $T_{follower}$ does not fit the lognormal distribution over the whole range lies in the nature of Digg.com. Tang \textit{et al.} \cite{Tang2011} showed that only a few users were active over a long period, while just a fraction of them was actually submitting stories. Only 2\% of all registered users succeeded to have their submissions \textquotedblleft promoted\textquotedblright\ to the frontpage \cite{Doerr-etal:2012}. Since the username of a submitter appears next to the story, these users receive a lot of followers during the relatively small period that their story was listed on the front page. Therefore, $f_{T_{follower}}(x)$ is large for small $x$ in Fig.~\ref{Fig_binsize1s_followers}. In summary, even ignoring the small time values, the random variable $T_{friend}$ and $T_{follower}$ possess different distributions. The process that generates $T_{friend}$ is more likely described by a variant of Barabasi's priority queue model, leading to power-law behavior. Since the random variable
$T_{follower}$ is generated by the collective dynamics of different individuals (that are likely weakly dependent), central limit law arguments may point towards the lognormal distribution \cite[p. 121-126]{PVM_PAComplexNetsCUP}.

Similar properties occur in sending and receiving emails. Obviously, a user can only send an email when he is online, whereas emails arrive at a user's inbox at his email server at all times. We analyzed the durations between receiving and sending emails in the Enron data set, which contains emails of all employees of Enron during roughly 6 years, starting in January 1998 until February 2004. Similar distributions arise as shown later in Figs.~\ref{Fig_enron_sent_diff} and \ref{Fig_enron_received_diff}.

\begin{figure}[ht]
\begin{center}
\includegraphics[width=1.00\linewidth]{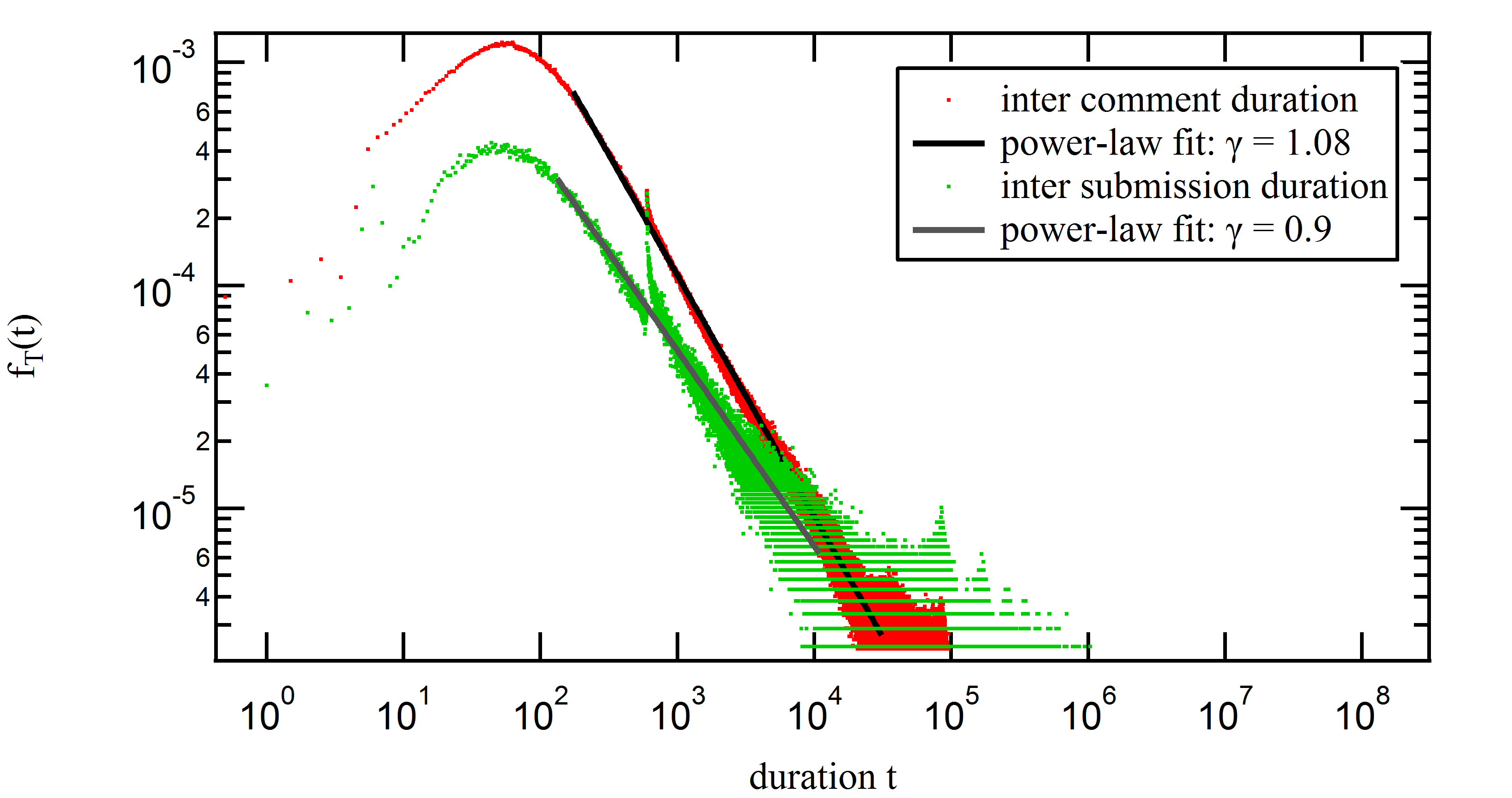}
\caption{Time difference between commenting and bookmark submission in Reddit.com}
\label{Fig_reddit_comment_and_submission}
\end{center}
\end{figure}

A third data set from Reddit.com\footnote{The Reddit.com data-set is hosted at Google BigQuery (bigquery.cloud.google.com/dataset/fh-bigquery).}, an OSN in which users mainly submit, comment or vote on bookmarks. The pdf of the duration $T$ between consecutive comments and submissions in Reddit are shown in Fig.~\ref{Fig_reddit_comment_and_submission}.

The tails of the distributions in Fig.~\ref{Fig_reddit_comment_and_submission} seem nicely fitted by a power-law distribution with exponents $\gamma\approx 1$. However, as mentioned in the introduction, the increasing regime in $f_{T}\left(  t\right)  $ for small values of $t$, the peak of $f_{T}\left( t\right) $ nor the concave form of the pdf can be modeled by a power-law distribution.

\section{Fitting a lognormal distribution}

\label{Sec_fittinglognormal}

\begin{figure}[ht]
\begin{center}
\includegraphics[width=1.00\linewidth]{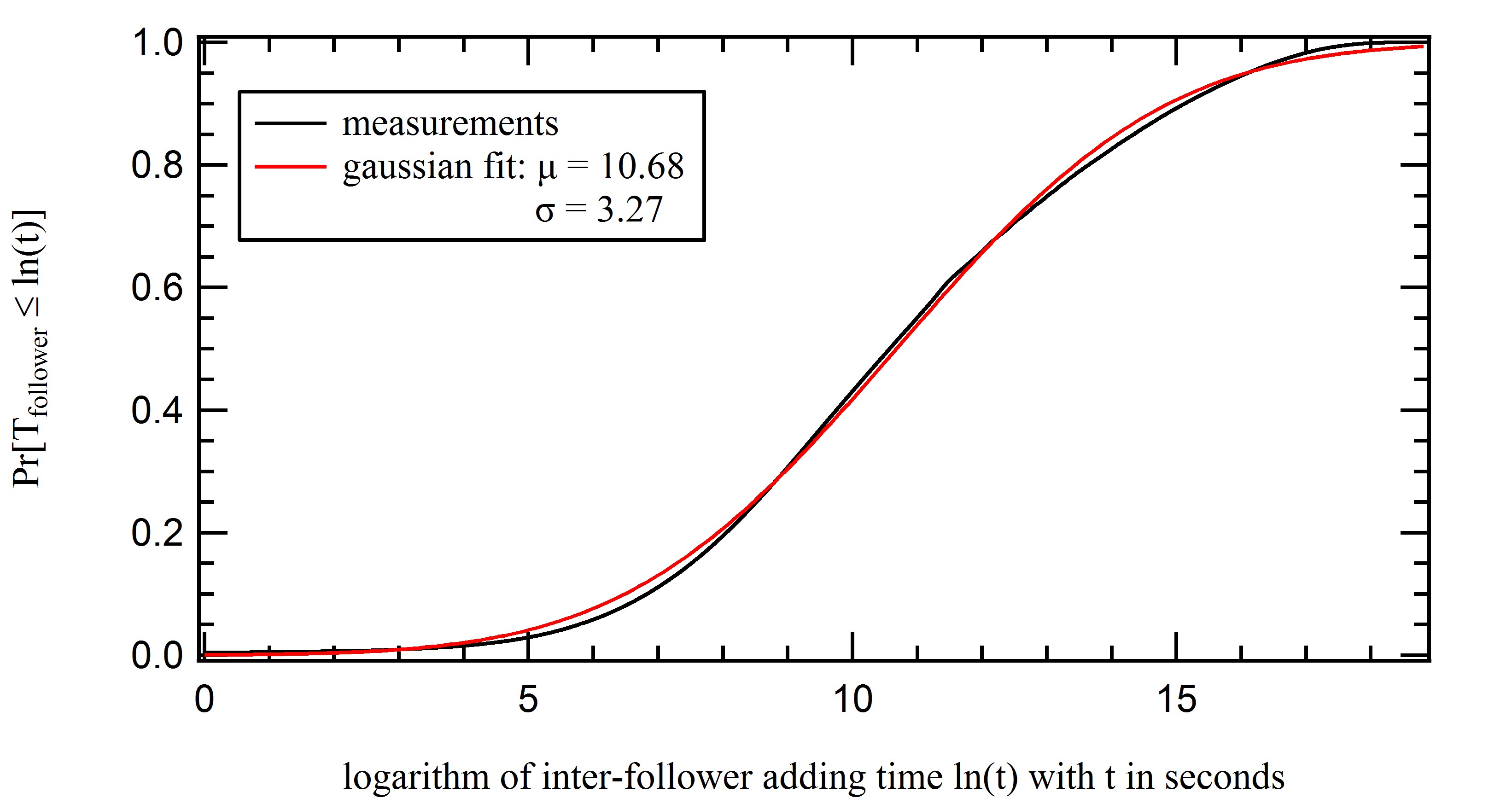}
\caption{Empirical distribution function (EDF) of the inter-follower durations $T_{follower}$ (same data as in Fig. \ref{Fig_binsize1s_followers}).}
\label{Fig_ccdf_seconds_normal_followers}%
\end{center}
\end{figure}

The two main approaches are based on the pdf and the EDF (empirical distribution function\footnote{The empirical distribution function is sometimes also called \emph{empirical cdf}: the cumulative distribution
function (cdf) based on an empirical measure.}), after a logarithmic transformation of the data. Figure~\ref{Fig_ccdf_seconds_normal_followers} depicts the data fitted\footnote{Fitting data to a EDF usually flattens interesting parts of a pdf, especially the tail of a distribution. Still, the benefit lies in the fact that binning is not needed and \textquotedblleft raw\textquotedblright\ data can be directly fitted \cite[p. 580-581]{PVM_PAComplexNetsCUP}.} to the distribution function of a normal distribution (\ref{distribution_lognormal}) and shows that the parameters of the EDF are about the same as those in Fig.~\ref{Fig_binsize1s_followers} (pdf approach).

\subsection{The effect of binning}

\label{Sec_effect_binning}

\subsubsection{Theory}

We generate \cite[p. 41]{PVM_PAComplexNetsCUP} $n$ realizations of a lognormal random variable $X$, which we denote by the set $\left\{  x_{k}\right\} _{1\leq k\leq n}=$ $\left\{  x_{1},x_{2},\ldots,x_{n}\right\} $. All $n$ realizations lie in the interval $\left[  x_{\min},x_{\max}\right]  $, where the minimum value is $x_{\min}=\min_{1\leq k\leq n}x_{k}$ and the maximum value is $x_{\max}=\max_{1\leq k\leq n}x_{k}$. The binning operation \cite[p. 580-581]{PVM_PAComplexNetsCUP} divides the entire data interval $\left[
x_{\min},x_{\max}\right]  $ into $m$ sub-intervals of length $\Delta x=\frac{x_{\max}-x_{\min}}{m}$ and the $j$-th subinterval $\left[  x_{\min }+\left(  j-1\right)  \Delta x,x_{\min}+j\Delta x\right]  $ for $1\leq j\leq m$ is associated with a bin $h_{j}$, that contains the number of realizations of $X$ or data points of set $\left\{  x_{k}\right\}  _{1\leq k\leq n}$ lying within the $j$-th subinterval,\[
h_{j}\approx n\int_{x_{\min}+\left(  j-1\right)  \Delta x}^{x_{\min}+j\Delta x}f_{X}\left(  u\right)  du \] Clearly, for a given set $\left\{  x_{k}\right\}  _{1\leq k\leq n}$, increasing the binsize $\Delta x$ decreases the number of bins $m$.

\begin{figure}[ht]
\begin{center}
\includegraphics[width=1.00\linewidth]{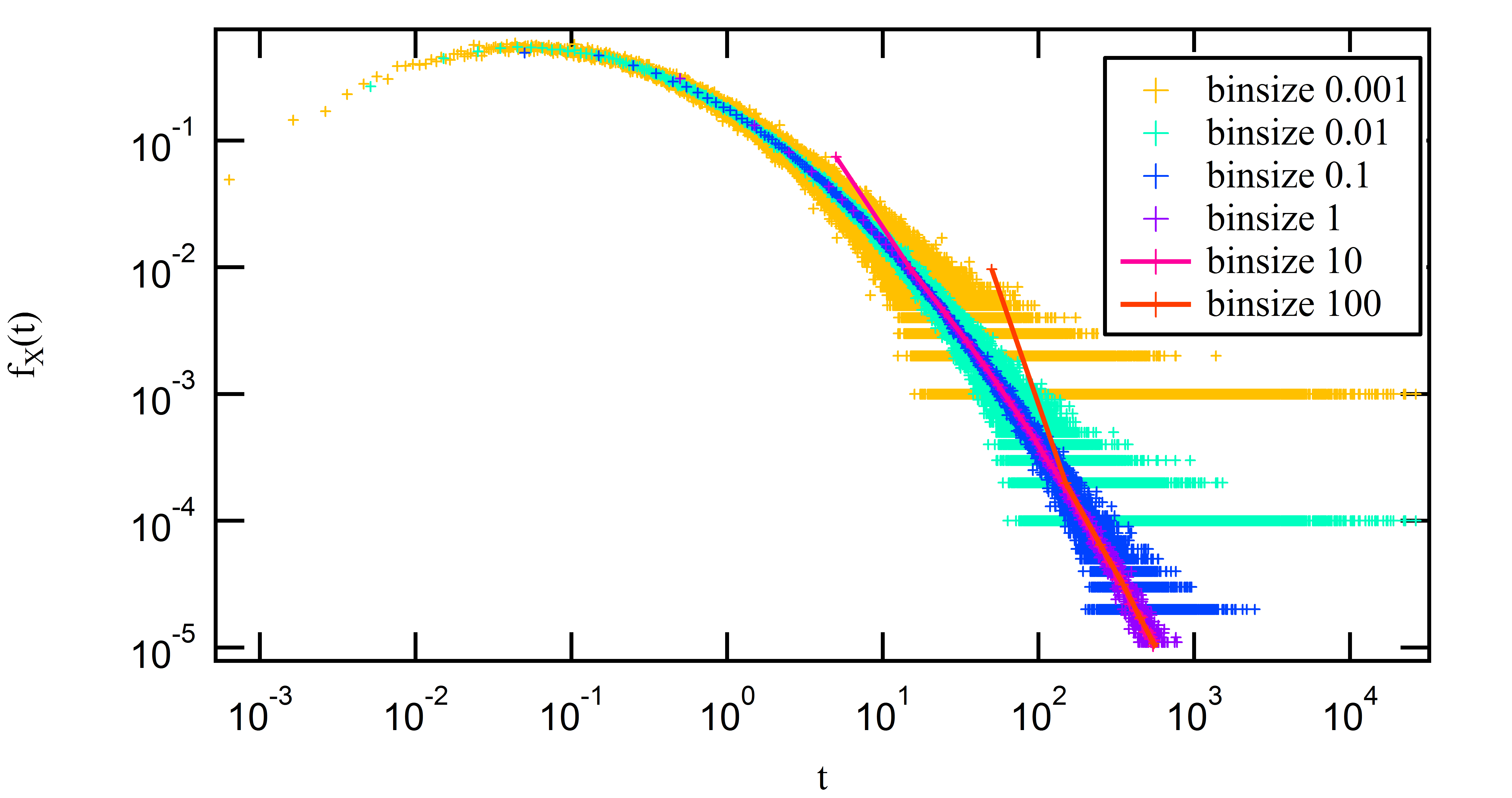}
\caption{The effect of different binsizes on a lognormal random variable}
\label{Fig_different_binning_m10_s2_lognormnoise}
\end{center}
\end{figure}

The effect of binning on the pdf $f_{X}\left(  t\right)  $ is depicted in Fig.~\ref{Fig_different_binning_m10_s2_lognormnoise}, where $n=10^{6}$ realizations from a lognormal random variable with parameters $\mu=10$ and $\sigma=2$ are drawn. As demonstrated in Appendix \ref{sec_properties_lognormal}, the scaled random variable $Y=bX$ has parameter $\mu_{Y}=\mu_{X}+\ln b$, but $\sigma_{Y}=\sigma_{X}$, implying that
\textquotedblleft binning\textquotedblright\ only changes the parameter $\mu$, but leaves the parameter $\sigma$ invariant! By binning (scaling) the distribution with different binsizes, the \textquotedblleft up-going regime\textquotedblright, where $f_{X}\left(  t\right)  $ increases with $t$, disappears and the observable part of the distribution \textquotedblleft evolves\textquotedblright\ towards a straight line on a log-log plot. Binning the data with larger binsizes decreases the parameter $\mu$, even to the extent that $\mu$ may become negative. If $\mu-\sigma^{2}$ decreases, the maximum of the lognormal at $e^{\mu-\sigma^{2}}$ tends to zero (infinitely far to the left on a log-log scale). Therefore, just the decreasing part of the quadratic shape of (\ref{pdf_lognormal}) will be visible in a log-log plot.

\subsubsection{Data analysis}

\begin{figure}[ptb]
\begin{center}
\includegraphics[width=1.00\linewidth]{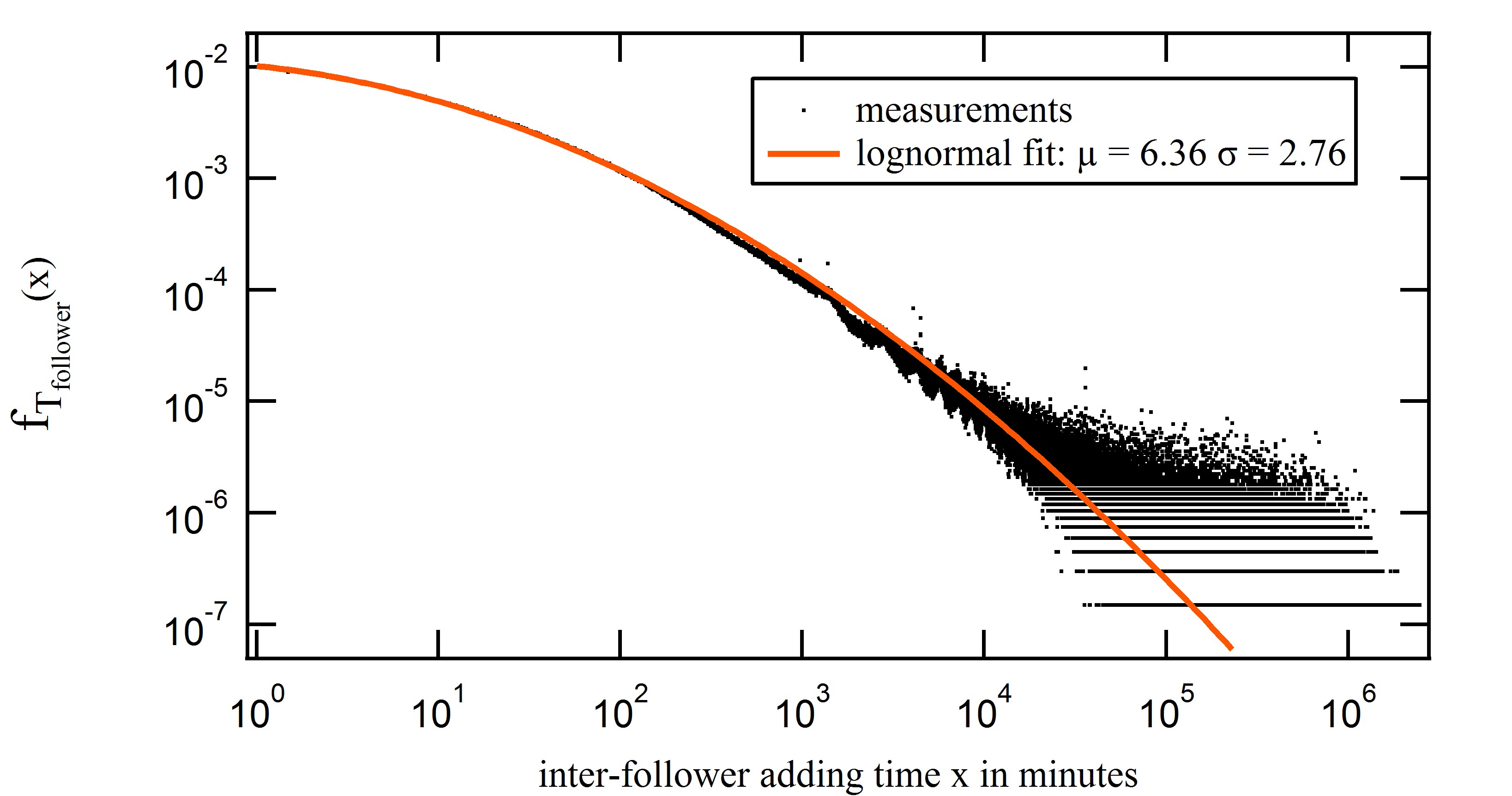}
\caption{The probability density function (pdf) of the inter follower times $T_{follower}$ binned per minute.}
\label{Fig_pdf_minutes_lognormal_followers}
\end{center}
\end{figure}

Binning the data in different time units, say per minute instead of per second, scales the distribution by a factor of 60 (since 60 seconds equals 1 minute), which will shift the parameter $\mu_{\min}$ of the CDF towards the left to $\mu_{\min}=\mu_{\sec}-\ln(60)\approx\mu_{\sec}-4.1$. However, a remarkable property of the lognormal distribution is that the parameter $\sigma$ will not change after linear scaling as shown in the Appendix \ref{sec_properties_lognormal}.
\begin{figure}[ptb]
\begin{center}
\includegraphics[width=1.00\linewidth]{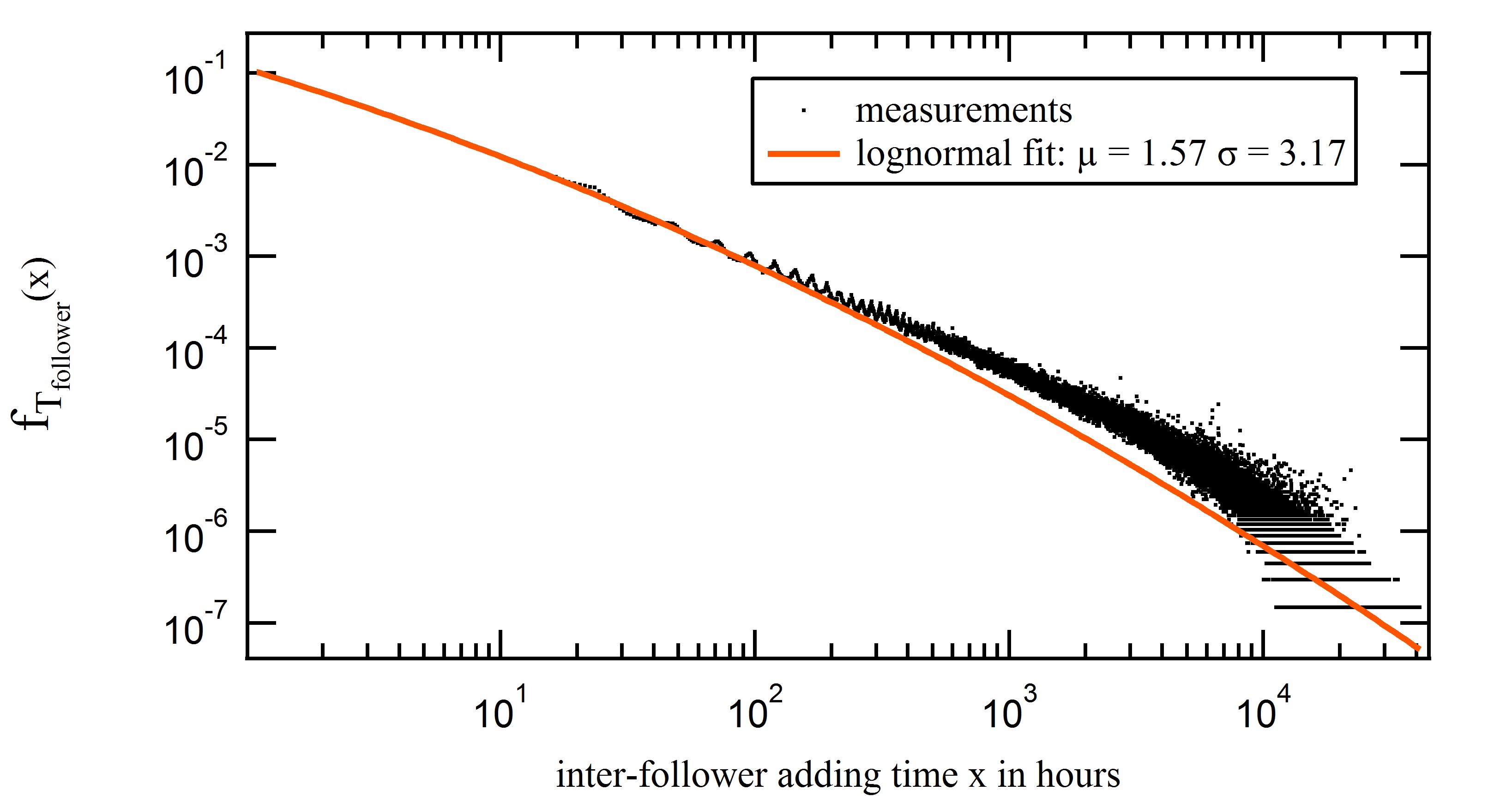}
\caption{The probability density function (pdf) of the inter follower times $T_{follower}$ binned per hour.}
\label{Fig_pdf_hours_lognormal_followers}
\end{center}
\end{figure}

\begin{figure}[ht]
\begin{center}
\includegraphics[width=1.00\linewidth]{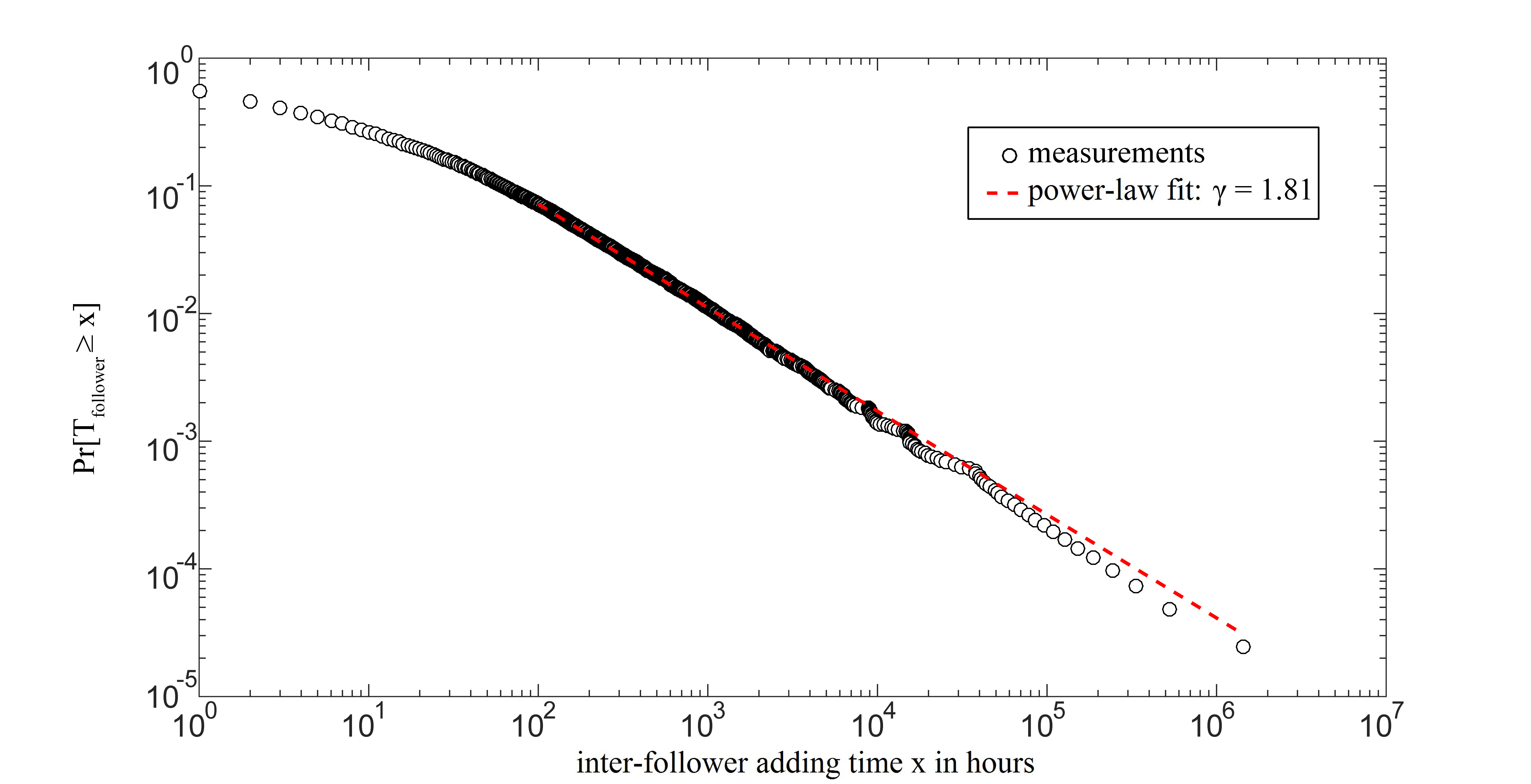}
\caption{The ccdf of the inter-follower time $T_{Follower}$ where data is binned per hour.}
\label{Fig_pdf_hours_lognormal_followers_exponential}
\end{center}
\end{figure}

An important consequence of a binning operation is illustrated in the pdfs in Fig.~\ref{Fig_pdf_minutes_lognormal_followers} and Fig.~\ref{Fig_pdf_hours_lognormal_followers}, that show the pdf of $T_{follower}$ binned per minute and per hour, respectively. Conforming to the theory, the parameter $\mu$ decreases by a factor of about $\ln60\approx4.1$, while the parameter $\sigma$ keeps its value, but the shape of the distribution changes. The distribution, binned per hour, is shown in Fig.~\ref{Fig_pdf_hours_lognormal_followers}, where the distribution can be confounded with a power-law distribution (with exponential cut-off). Moreover, this misinterpretation may be justified, because the exponent $\gamma=1.57$ of a power-law fitted to the hourly binned data is close to the exponent $\gamma=1.53$ found for the $T_{friend}$ distribution in Fig. \ref{Fig_interfriendaddtime_seconds}. Fig. \ref{Fig_pdf_hours_lognormal_followers_exponential} shows the complementary cumulative distribution function of $T_{follower}$ binned per hour, fitted using the method of Clauset \textit{et al.} \cite{Clauset:2009}. The resulting power-law exponent in Fig. \ref{Fig_pdf_hours_lognormal_followers_exponential} is $\gamma=1.81$. The visualization in Fig. \ref{Fig_pdf_hours_lognormal_followers_exponential} is misleading because the original distribution is, as shown earlier, a lognormal distribution.

For completeness we like to refer to Table \ref{tab:EstimatedParametersUsingTheMethodOfClausetEtAl} in Appendix \ref{sec_likelihood_testing}, where we list fitted parameters, p values and the result of log-likelihood tests, conducted using the method of Virkar and Clauset \cite{virkar2014}. As shown in the table, the larger the used binsizes, the higher the p values indicating that a power-law fits the data. For small binsizes (low p values) the results are not conclusive or rather in favor of a lognormal distribution.

To further demonstrate the effect, Figs.~\ref{Fig_pdf_minutes_lognormal_followers}, \ref{Fig_pdf_hours_lognormal_followers} and \ref{Fig_pdf_hours_lognormal_followers_exponential} depict distributions in which the data was artificially binned by minutes and hours. In reality, such binning operation might occur if the data provider (the OSN or web service) returns values per hour or even larger scales. In other words, data measured in larger time steps over a certain total time range thus erases the possibility to distinguish between a lognormal and power-law distribution.
\begin{figure}[ht]
\begin{center}
\includegraphics[width=1.00\linewidth]{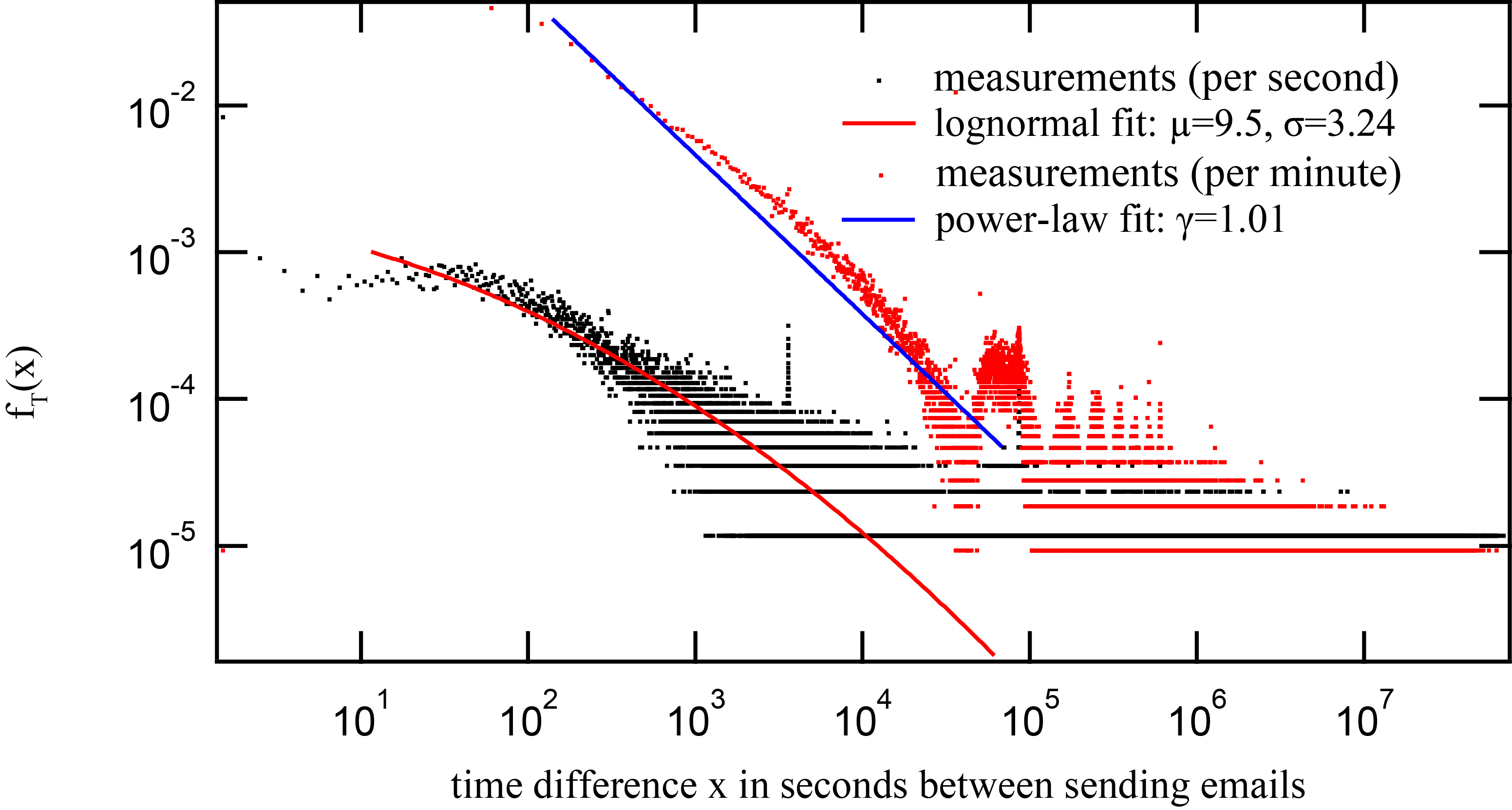}
\caption{Time difference between consecutively sent emails per person in the Enron data set.}
\label{Fig_enron_sent_diff}
\end{center}
\end{figure}

The sending time of an email in the Enron data set until May 2001 was stored per minute, whereas afterwards (June 2001 - February 2004) the sending time is stored with an accuracy of a second. The resulting pdf of time differences between sent emails is plotted in Fig.~\ref{Fig_enron_sent_diff}, where the black and red dots represent the measurements per second and per minute, respectively. The parameters of the fitted lognormal ($\mu=9.5,\sigma=3.24$) and power-law distribution $(\gamma\approx1$) correspond to those of Digg.com in Fig. \ref{Fig_binsize1s_followers} and Fig. \ref{Fig_interfriendaddtime_seconds}.
\begin{figure}[ht]
\begin{center}
\includegraphics[width=1.00\linewidth]{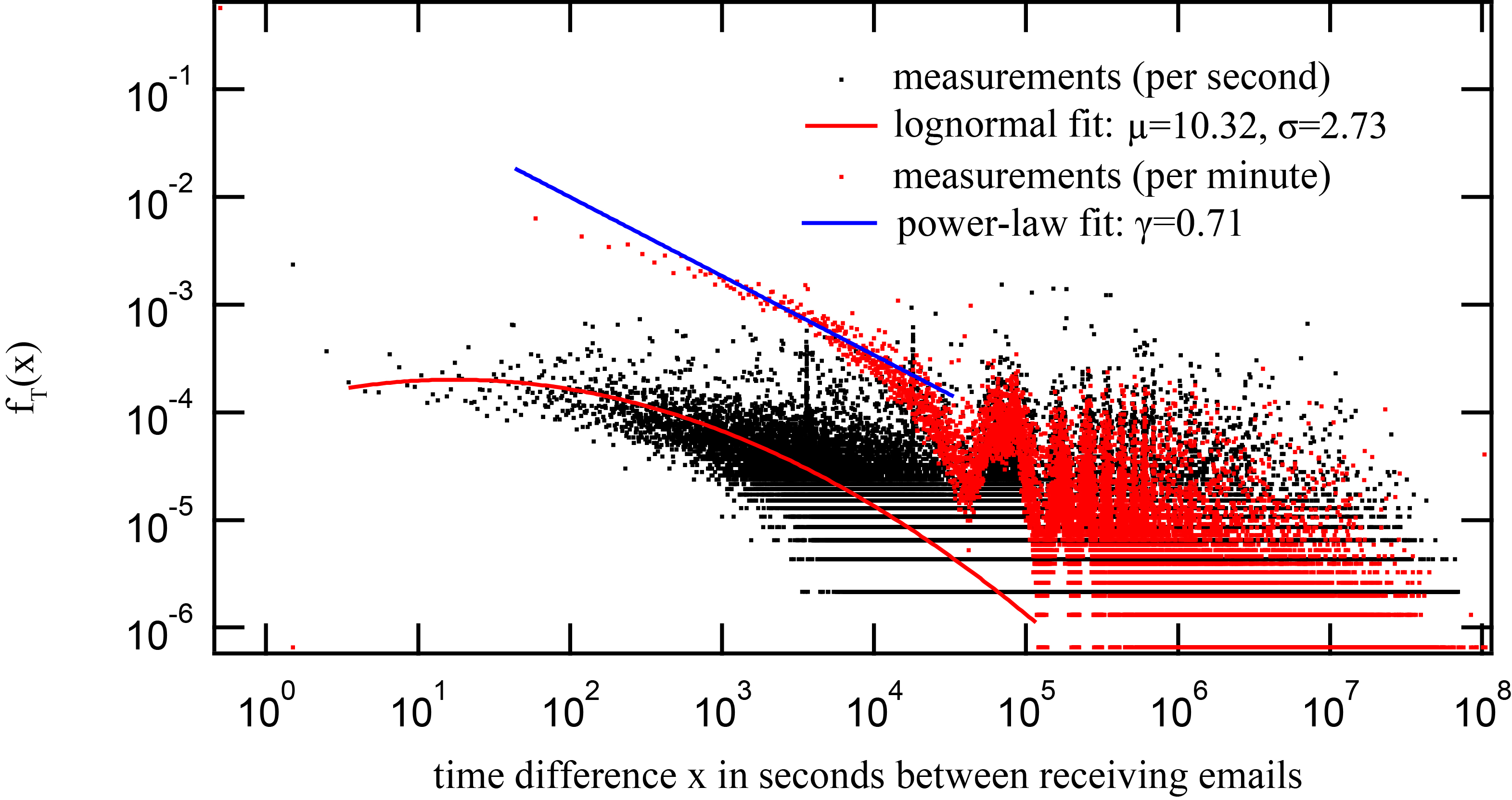}
\caption{Time difference between consecutively received emails per person in the Enron data set.}
\label{Fig_enron_received_diff}
\end{center}
\end{figure}

The pdf of inter-arrival times of received emails, split in observations per minute (red) and per second (black) in Fig.~\ref{Fig_enron_received_diff}, can be compared to the durations of added followers, where the parameters are again in the same range as shown before. We found that the EDF of the data (not shown) agrees with the pdfs in Fig.~\ref{Fig_enron_sent_diff} and Fig.~\ref{Fig_enron_received_diff}, underlining the likeliness that a lognormal random variable models the data better than a power-law.

\section{Reported parameters of power-law and lognormal distributions}

\label{Sec_reported_parameters} Table \ref{tab:PowerLawExponentsFoundInHumanDynamics} lists inter-arrival times and the fitted exponents found in publications in which the aspect of binning was ignored. Rather small power-law exponents $\gamma$ between $0.7$ and $1.8$ are found in multiple data sets of human activity.

\begin{table}[th]
\begin{center}
\begin{footnotesize}
\begin{tabular}
[c]{lll}
Data-set & Exponent ($\gamma$) & Reference\\\hline
Time between sending Emails & 1 & Barabasi (2005) \cite{Barabasi05nature}\\
Time between sending Emails & 1.2 & Eckmann \textit{et al.} (2004) \cite{Eckmann2004}\\
Time between clicks in website usage & 1, 1.25 & Gon\c{c}alves and Ramasco (2008) \cite{goncalves2008}\\
Time between similar actions in web data & 1.1, 1.2, 1.8 & Radicchi (2008) \cite{Radicchi2008a}\\
Time between messages in instant-messaging & 1.53 & Leskovec and Horvitz (2008) \cite{Leskovec_msn2008}\\
Time between phone calls & 0.9 & Candia \textit{et al.} (2008) \cite{Candia2008}\\
Time between phone calls & 0.7 & Karsai \textit{et al.} (2012) \cite{Karsai_NSR2012}\\
Time between sending Emails & 1 & Karsai \textit{et al.} (2012) \cite{Karsai_NSR2012}\\
Time between sending short messages & 0.7 & Karsai \textit{et al.} (2012) \cite{Karsai_NSR2012}%
\end{tabular}
\end{footnotesize}
\end{center}
\caption{Power-law exponents found for durations between technology related human dynamics}%
\label{tab:PowerLawExponentsFoundInHumanDynamics}%
\end{table}

Lognormal distributions were reported for similar human activity as shown in Table \ref{tab:addlabel}, which extends the collection of Limpert \textit{et al.} \cite{Limpert:2001}. The oldest analysis was conducted by Boag \cite{Boag49} in 1949. Based on the scaling invariance of $\sigma$ as demonstrated in Appendix \ref{sec_properties_lognormal}, the parameter $\sigma$ can be compared over different measurements, whereas the parameter $\mu$ cannot, since $\mu$ depends on the units in which the lognormal random variable is measured. Interestingly, all $\sigma$'s in Table \ref{tab:addlabel} lie within an amazingly small range of $0.35\leq\sigma \leq3.2$, which shows that the parameter $\sigma$ only varies over about one order of magnitude in different measured phenomena. If $\sigma$ is rather small as in our measurements where $2.73\leq\sigma\leq3.24$, the first term in (\ref{loglogform}) dominates and a quadratic function appears in a log-log plot. Consequently, these rather small values of $\sigma$ and those in Table \ref{tab:addlabel} contradict the common deductions made from (\ref{loglogform}) in Appendix \ref{sec_properties_lognormal}, namely that only for large values of $\sigma$ power-law and lognormal distributions are indistinguishable. Equations (\ref{pdf_lognormal_power_law_like_form}) and
(\ref{power_law_exponent_lognormal}) in Appendix \ref{sec_properties_lognormal} explain why the lognormal distribution may seem linear in a certain regime.

\begin{table}[ptbh]
\caption{Literature of lognormal distributions (excerpt)}
\label{tab:addlabel}
\begin{center}
\begin{footnotesize}
\begin{tabular}
[c]{llll}
$\mu$ & $\sigma$ & process & reference\\\hline
5.547 & 2.126 & email forwarding & Iribarren and Moro \cite{Iribarren09}\\
$\approx8$ & $\approx2$ & email forwarding & Stouffer \textit{et al.} \cite{Stouffer:2005vn}\\
$\mu_{1} = 1$ hour &  &  & \\
$\mu_{2} = 2$ days &  & email forwarding & Stouffer \textit{et al.} \cite{stouffer06}\\
2.47 & 0.38 & infection times & Nishiura \cite{Nishiura07}\\
2.47 & 0.36 & latency periods of diseases & Limpert \cite{Limpert:2001}\\
14 days & 1.14 & latency periods of diseases & Sartwell \cite{Sartwell:1950}\\
100 days & 1.24 & latency periods of diseases & Sartwell \cite{Sartwell:1950}\\
2.3 hours & 1.48 & latency periods of diseases & Sartwell \cite{Sartwell:1950}\\
2.4 days & 1.47 & latency periods of diseases & Sartwell \cite{Sartwell:1950}\\
12.6 days & 1.50 & latency periods of diseases & Sartwell \cite{Sartwell:1950}\\
21.4 days & 2.11 & latency periods of diseases & Sartwell \cite{Sartwell:1950}\\
9.6 months & 2.5 & survival times after cancer diagnosis & Boag \cite{Boag49}\\
15.9 monts & 2.8 & survival times after cancer diagnosis & Feinleib and Macmahon \cite{feinleib60}\\
17.2 months & 3.21 & survival times after cancer diagnosis & Feinleib and Macmahon \cite{feinleib60}\\
14.5 months & 3.02 & survival times after cancer diagnosis & Boag \cite{Boag49}\\
60 years & 1.16 & age of onset of Alzheimer & Horner \cite{Horner87}\\
4 days &  & incubation periods (viral infections) & Lessler \textit{et al.} \cite{Lessler09}\\
3 to 5 & $\approx2$ & task completion & Linden \cite{Linden:2006}\\
0.5 & 1.4 & strike duration & Lawrence \cite{Lawrence:1984}\\
&  & time of individual activities & Mohana \textit{et al.} \cite{mohana2006}\\
0.43 & 1.634 & call duration & Spedalieri \textit{et al.} \cite{SpedalieriMB05}\\
3.5 & 0.70 & message holding time & Barcelo and Jord\'{a}n \cite{Barcelo99}\\
7.439 & 0.846 & transmission holding time & Barcelo and Jord\'{a}n \cite{Barcelo99}\\
3.29 & 0.890 & channel holding time & Barcelo and Jord\'{a}n \cite{Barcelo99}\\
3.3 & 0.89 & channel holding time & Barcelo and Jord\'{a}n \cite{Barcelo99}\\
$\mu_{1}=1.31$ & $\sigma_{1}=0.32$ &  & \\
$\mu_{2}=2.26$ & $\sigma_{2}=0.56$ & call holding time & Bolotin \cite{Bolotin94}\\
&  & citations & Eom and Fortunato \cite{Eom11}\\
&  & citations & Redner \cite{Redner05}\\
1 to 2 & $0.35-0.45$ & citations & Stringer \textit{et al.}\cite{Stringer08}\\
& 1.095 & citations & Radicchi \textit{et al.}\cite{Radicchi:2008}\\
$\mu_{1}=3.7$ & $\sigma_{1}=0.8$ &  & \\
$\mu_{2}=5.6$ & $\sigma_{2}=3.1$ & retweeting behavior & Doerr \textit{et al.}\cite{doerr2013-infection}\\
5.29 & 0.42 & distribution of votes on pages of Digg.com & Van Mieghem \textit{et al.} \cite{Mieghem_lognormal_2011}\\
&  &  &
\end{tabular}
\end{footnotesize}
\end{center}
\end{table}

\section{The Historical Debate of the Power-law versus Lognormal Distribution}

\label{Sec_debate}

Mitchell \cite{Mitchell_boek2009} discusses the fascinating and abundant appearance of power-law distributions and mentions processes such as self-criticality, that are related to phase transitions, as the main producers of power-laws, but she remarks that the precise nature as well as the deviations from a power-law are still open to debate. Table \ref{tab:addlabel} illustrates that the research on lognormal distributions has a long history. Unfortunately, also the processes that produce (almost) lognormal behavior are not well understood. Mitzenmacher \cite{mitzenmacher03} overviews processes leading to power-law and lognormal distributions, emphasizing that minor changes in the process will lead to either a power-law or a lognormal distribution. Mitzenmacher mentions the work of Gabaix \cite{Gabaix99}, who analyzed the size distribution of cities in the United States. Interestingly, Gabaix \cite{Gabaix99} found that the city size distribution follows a Zipf distribution, which is a power-law distribution (\ref{pdf_powerlaw}) with an exponent $\gamma=1$. Gabaix argues that cities cannot become infinitely small, a fact that imposes a lower bound to their size. When modeling the size of cities as a Markov chain with a fixed number of cities, which grow stochastically as proposed by Gibrat \cite{Gibrat:1931}, then the steady state
of the Markov chain will follow Zipf's distribution with an exponent $\gamma=1$. If there is no lower bound on the city size, so that cities can be arbitrarily small, then the distribution degenerates to a lognormal distribution.

Gibrat \cite{Gibrat:1931}, whose work is often associated with the law of proportionate effect \cite{Mieghem_lognormal_2011}, argues in his model that transition probabilities or the variance of transition probabilities in a growth process are independent from the size. Gibrat, who estimated the distribution of city sizes, concluded that the law of proportionate growth leads to a distribution that is lognormal. However, Simon \cite{Simon1955} showed that Gibrat's law of the proportionate effect may lead to other
heavy-tailed distributions as well.

Champernowne \cite{Champernowne53} and Cordoba \cite{Cordoba08} analyzed the income distribution of England and Wales with Markov theory. Again, their crucial assumption is that incomes have a lower bound.

Eeckhout \cite{Eeckhout04} analyzed the distribution of city sizes by using accurate data from the US census in 2000 and found that the heavy tail obeys Zipf's law. But, the entire distribution is better described by a lognormal than a Pareto distribution. By comparing the census data from 1990 and 2000,
Eeckhout shows that the growth of a city is independent of its size. The parameters of the lognormal distribution found by Eeckhout are $\mu=7.28$ and $\sigma=1.75$.

The difference with the above mentioned Markov chain approach lies in the fact that Eeckhout \cite{Eeckhout04} modeled the process by a multiplicative process, which leads to a lognormal distribution. This multiplicative process, proposed by Kapteyn \cite{Kapteyn18} in 1903 for the first time and later
coined the \textquotedblleft Law of Proportionate Effect\textquotedblright\ by Gibrat \cite{Gibrat:1931}, is based on the central limit theorem applied to a multiplicative process, which leads to lognormal distributed sizes \cite{PVM_PAComplexNetsCUP}. Gibrat's growth process is defined as
\begin{equation}
S_{t}=a_{t}\times S_{t-1} \label{Gibrat_governing_eq}%
\end{equation}
where the size $S_{t}$ of an item at state $t$ depends upon the previous size $S_{t-1}$ times a positive, random factor $a_{t}$. By taking the logarithm of both sides in (\ref{Gibrat_governing_eq}) and denoting $\xi_{t}\equiv\ln a_{t}$, we obtain, after iteration, \[\ln S_{t}=\ln S_{0,i}+\xi_{1}+\xi_{2}+\ldots+\xi_{t}
 \] By the Central Limit Theorem \cite{PVM_PAComplexNetsCUP}, $\frac{\sum_{k=1}^{t}\xi_{k}-t\mu}{\sigma\sqrt{t}}\overset{d}{\rightarrow}N(0,1)$, we arrive at, for large $t$,\[ \Pr[\ln S_{t}\leq y]\rightarrow\frac{1}{\sqrt{2\pi}}\int_{-\infty}^{\frac{y-t\mu}{\sigma\sqrt{t}}}e^{-u^{2}/2}\;du \] from which approximately \[ \ln S_{t}\cong t\mu+\sqrt{t}\sigma N(0,1) \] where $\mu$ denotes the mean and $\sigma^{2}$ the variance of the sequence $\{\xi_{k}\}_{1\leq k\leq t}$. 

For large $t$, a power-law distribution (\ref{pdf_powerlaw}) tends to zero as $O\left(  t^{-\gamma}\right)  $, while a lognormal distribution (\ref{pdf_lognormal}) tends considerably faster as $O\left(  t^{-1}\exp\left[ -\frac{\log^{2}t}{2\sigma^{2}}\right]  \right)  $ to zero, illustrating that the deep tails are significantly different. Malevergne \cite{Malevergne09} addressed this fact by the concept of a slowly varying function, in particular, for $x\rightarrow\infty$ and $t>0$, the power-law distribution
(\ref{pdf_powerlaw}) features \[ \lim_{x\rightarrow\infty}\frac{f_{X}(t\cdot x)}{f_{X}(x)}=t^{-\gamma} \]
The lognormal distribution (\ref{pdf_lognormal}) on the other hand is not slowly varying. In the limit $x\rightarrow\infty$, a lognormal distribution will, for $t>1$, always tend to zero: \[ \lim_{x\rightarrow\infty}\frac{f_{X}(t\cdot x)}{f_{X}(x)}=\lim_{x\rightarrow \infty}\frac{1}{t}e^{-\frac{(\ln(t))^{2}}{2\sigma^{2}}}e^{-\ln(t)\cdot \frac{\ln(x)-\mu}{\sigma^{2}}}=0 \] This different tail behavior questions whether exponential bin sizes should be used in a log-log plot, because these may hide the rapid decrease in the tail. 

\section{An explanation of lognormal human interactivity times}

\label{sec_plausible_explanation_lognormal_interactivity_times}
Perhaps, a plausible explanation of the appearance of lognormal human interactivity times is that the logarithm $\log T$ of a human interactivity time $T$ is Gaussian or normally distributed (see Appendix \ref{sec_properties_lognormal}), most likely as a consequence of the Central Limit Theorem. Roughly speaking, the Central Limit Theorem applies for a large number of weakly dependent random variables, where none of them is dominant. Thus, rather than concentrating on the time $T$, it seems more natural to focus on the random variable $\log T$ as the decisive quantity. The logarithm of a human related measure often occurs: for example, in our hearing system, the intensity of sound is logarithmically experienced. But more importantly, the large variability in human performance also seems logarithmically distributed
\cite{Shockley_1957,Linden:2006} and the measured interactivity times are strongly related to the large differences in human performance and/or behavior. 

On the other hand, if $Y$ is an exponential random variable with mean $\frac{1}{\gamma-1}$ for $Y\geq\log\tau$ (else $Y=0$) and $X=e^{Y}$, then \[ \Pr\left[  X\leq t\right]  =\Pr\left[  Y\leq\log t\right]  =1-e^{-\left( \gamma-1\right)  \left(  \log t-\log\tau\right)  }=1-\left(  \frac{t}{\tau }\right)  ^{1-\gamma} \] so that $\frac{d}{dt}\Pr\left[  X\leq t\right]  =\frac{\gamma-1} {\tau^{1-\gamma}}t^{-\gamma}$ is a perfect power-law probability density function as in (\ref{pdf_powerlaw}). However, if $\log T$ $\geq\log\tau$ were exponentially distributed with mean $\frac{1}{\gamma-1}$, the memoryless property of the exponential distribution \cite[p. 43]{PVM_PAComplexNetsCUP} would indicate that \[ \Pr\left[  \log T\geq t+u|\log T>u\right]  =\Pr\left[  \log T\geq t\right] \] In other words, given that the logarithm of a human interactivity time is larger than $u$ time units, the probability that $\log T$ exceeds $t+u$ time units is equal to the probability that $\log T$ exceeds $t$ time units, for any $u$ and $t$ larger than $\log\tau$, precisely as if the log-interactivity time $u$ never has been spent or waited, which is quite counterintuitive for a duration between consecutive activities. Alternatively, with $u=\log s$ and $s>\tau$, the memoryless property of $\log T$ implies \textquotedblleft scale-freeness in $T$\textquotedblright: \[ \Pr\left[  \log\frac{T}{s}\geq t|\log\frac{T}{s}>0\right]  =\Pr\left[  \log T\geq t\right] \] which shows, ignoring the condition $\log\frac{T}{s}>0$, the independence on the \textquotedblleft scale $s$\textquotedblright. But, we have shown in Section \ref{Sec_effect_binning} that rescaling the human interactivity time (by different bin sizes) definitely alters the distribution.

From these arguments, we can infer that a power-law time $T$ is less defendable than a lognormally distributed $T$.

\section{Conclusion}

\label{sec_conclusion} In this paper, the interactivity durations of individuals, between creating friendship relations, writing emails, commenting and voting on online content are analyzed. We found that the distribution of durations to add friends follows a power-law with an exponent of $\gamma \simeq1.8$, whereas the durations to acquire followers are well described by a lognormal with $\mu\approx10.5$ and $\sigma\approx2.8$. Due to the small probability of executing two tasks in a small time interval (typically ignored in fitting a power law), we claim that a lognormal distribution covers the entire activity time range better than a power distribution.

In addition, we show that binning of lognormally distributed data can seriously affect the perception: the parameter $\mu$ shifts towards smaller values, but the parameter $\sigma$ of a lognormal distribution does not change after a binning or scaling operation. In the extreme case, only the heavy tail of the lognormal distribution (\ref{pdf_lognormal}) remains, which follows a power-law distribution (\ref{pdf_powerlaw}) with an exponent of $\gamma$ close to 1. As explained in the Appendix, there exists an interval in which the lognormal distribution is indistinguishable from a power-law distributions with power-law exponent $\gamma\approx1+\varepsilon$, for small $\varepsilon >0$.

Similar observations and concerns, discussed for a long time in the literature of city size and income distributions and reviewed in Section \ref{Sec_debate}, supports that a lognormal distribution is modeling the whole data range better than power-laws. Finally, Section \ref{sec_plausible_explanation_lognormal_interactivity_times} argues that the logarithm of a quantity associated with human activities rather than the quantity itself is the better descriptor, because human performance seems to fit a lognormal.

\appendix

\section{The lognormal random variable and distribution}

\label{sec_properties_lognormal}

A lognormal random variable \cite[p. 57]{PVM_PAComplexNetsCUP} is defined as $X=e^{Y}$ where $Y=N\left(  \mu,\sigma^{2}\right)  $ is a Gaussian or normal random variable. Hence, $X\geq0$. The distribution function $F_{X}(t)=\Pr[X\leq t]=\Pr[Y\leq\log t]$ is
\begin{equation}
F_{X}(t)=\frac{1}{\sigma\sqrt{2\pi}}\int_{-\infty}^{\log t}\exp\left[
-\frac{(u-\mu)^{2}}{2\sigma^{2}}\right]  du=\frac{1}{2}\left(  1+\text{erf}%
\left(  \frac{t-\mu}{\sigma\sqrt{2}}\right)  \right)
\label{distribution_lognormal}%
\end{equation}
where erf($x$) is the error function. The probability density function (pdf) of a lognormal random variable $X$ follows from the definition $f_{X}(t)=\frac{d}{dt}\Pr[X\leq t]$, for $t\geq0$, as (\ref{pdf_lognormal}), where $\left(  \mu,\sigma\right)  $ are called the parameters of the lognormal pdf, while the mean and variance are \cite[p. 57]{PVM_PAComplexNetsCUP} \[ E\left[  X\right]  =e^{\mu}e^{\frac{\sigma^{2}}{2}} \] and \[ \text{Var}\left[  X\right]  =e^{2\mu}e^{\sigma^{2}}\left(  e^{\sigma^{2} }-1\right) \] The limit $\sigma\rightarrow0$ reduces to a Dirac delta function at $t=e^{\mu}$, thus $\lim_{\sigma\rightarrow0}f_{X}\left(  t\right)  =\delta\left( t-e^{\mu}\right)  $.

Given the mean and variance, the parameters of the lognormal are found as 
\begin{equation}
\sigma^{2}=\log\left(  1+\frac{\text{Var}\left[  X\right]  }{\left(  E\left[ X\right]  \right)  ^{2}}\right)  \label{parameter_sigma^2}%
\end{equation}
and
\begin{equation}
\mu=\log E\left[  X\right]  -\frac{\sigma^{2}}{2} \label{parameter_mu}%
\end{equation}
Although $E\left[  X\right]  \geq0$, we remark that the parameter $\mu$ can be negative. Moreover, (\ref{parameter_sigma^2}) and (\ref{parameter_mu}) show that the scaled lognormal random variable $Y=bX$, where $b$ is a positive real number, has mean $\sigma_{Y}=\sigma$ and $\mu_{Y}=\mu+\log b$. Hence, \emph{scaling} by a factor $b$ does not change the parameter $\sigma$, which has interesting consequences for binning and measured data: the unit (e.g. second versus hours) in which the random variable is measured does not alter the parameter $\sigma$, only the parameter $\mu$.

The change of the argument $t\rightarrow e^{u}$ in $f_{X}\left(  t\right)  $ leads to
\begin{equation}
f_{X}\left(  e^{u}\right)  =e^{-\mu+\frac{\sigma^{2}}{2}}\frac{\exp\left[
-\frac{\left(  u-\left(  \mu-\sigma^{2}\right)  \right)  ^{2}}{2\sigma^{2}%
}\right]  }{\sigma\sqrt{2\pi}} \label{pdf_lognormal_e^u}%
\end{equation}
illustrating that the scaled lognormal pdf $e^{\mu-\frac{\sigma^{2}}{2}} f_{X}\left(  e^{u}\right)  $ is a Gaussian pdf $N\left(  \mu^{\prime},\sigma^{2}\right)  $ with mean $\mu^{\prime}=\mu-\sigma^{2}$. The maximum of $f_{X}\left(  t\right)  $ occurs at $t=e^{\mu-\sigma^{2}}$ and equals $\max_{t\geq0}f_{X}\left(  t\right)  =\frac{e^{-\mu}e^{\frac{\sigma^{2}}{2}}}{\sigma\sqrt{2\pi}}$, which follows directly from (\ref{pdf_lognormal_e^u}).
Moreover, we find easier from (\ref{pdf_lognormal_e^u}) than from (\ref{pdf_lognormal}) that $\lim_{u\rightarrow-\infty}f_{X}\left(e^{u}\right)  =f_{X}\left(  0\right)  =0$ and that $f_{X}^{\prime}\left(
0\right)  =0$. This means that any lognormal starts at $t=0$ from zero, increases up to the maximum at $t=e^{\mu-\sigma^{2}}>0$ after which it decreases towards zero at $t\rightarrow\infty$. Thus, the lognormal is
bell-shaped, but, in contrast to the Gaussian, the lognormal pdf is not symmetric around its maximum at $t=e^{\mu-\sigma^{2}}$ and can be seriously skewed.

The expression for the lognormal pdf in (\ref{pdf_lognormal}) can be rewritten \cite{Malevergne09} in a \textquotedblleft power-law\textquotedblright-like form as
\begin{equation}
f_{X}\left(  t\right)  =\frac{e^{-\frac{\mu^{2}}{2\sigma^{2}}}}{\sigma
\sqrt{2\pi}}t^{-\alpha\left(  t\right)  }
\label{pdf_lognormal_power_law_like_form}%
\end{equation}
where the exponent $\alpha\left(  t\right)  $ equals%
\begin{equation}
\alpha\left(  t\right)  =1+\frac{\log t-2\mu}{2\sigma^{2}}
\label{power_law_exponent_lognormal}%
\end{equation}
which illustrates that a lognormal random variable behaves as a power-law random variable, provided the last fraction in (\ref{power_law_exponent_lognormal}) is negligibly small, say $\varepsilon$.
The latter happens when $\left\vert \frac{\log t-2\mu}{2\sigma^{2}}\right\vert <\varepsilon$. Thus, when $t\in\lbrack e^{2\mu-2\sigma^{2}\varepsilon},e^{2\mu+2\sigma^{2}\varepsilon}]$ or in terms of the lower bound $\tau=e^{2\mu-2\sigma^{2}\varepsilon}$ and upper bound $\kappa=e^{2\mu+2\sigma^{2}\varepsilon}$ defined in Section \ref{sec_introduction}, the pdf (\ref{pdf_lognormal}) of a lognormal random variable is almost indistinguishable from the pdf (\ref{pdf_powerlaw}) of a power-law random variable with power exponent $\gamma\approx1+\varepsilon$. The $t$-interval $[e^{2\mu-2\sigma^{2}\varepsilon},e^{2\mu+2\sigma^{2}\varepsilon}]$ exceeds the maximum $e^{\mu-\sigma^{2}}$ of the lognormal pdf and is clearly longer
when $\sigma$ is larger (as well as the tolerated accuracy $\varepsilon$ increases). Malvergne \textit{et al.} \cite{Malevergne09} mention that, if $\sigma=3.4$, then the exponent $\alpha\left(  t\right)  $ in
(\ref{power_law_exponent_lognormal}) varies less than 0.3 units over a range of three orders of magnitude.

Taking the logarithm on both sides of (\ref{pdf_lognormal}) results in 
\begin{equation}
\ln(f_{X}(t))=-\frac{1}{2\sigma^{2}}\ln(t)^{2}+(\frac{\mu}{\sigma^{2}}%
-1)\ln(t)-\ln(\sqrt{2\pi}\sigma)-\frac{\mu^{2}}{2\sigma^{2}}
\label{loglogform}%
\end{equation}
If $\sigma$ is large, then the second term $(\frac{\mu}{\sigma^{2}}-1)\ln(t)$ in (\ref{loglogform}) dominates, which leads to a straight line with in a log-log plot resembling a power-law with exponent $\gamma=1$. On the other hand, if $\sigma$ is small or $\mu=\sigma^{2}$, the first, quadratic term in
(\ref{loglogform}) dominates.

\section{Distinguishing power-law from lognormals by likelihood testing}
\label{sec_likelihood_testing}

We apply the method of Clauset \textit{et al.} \cite{Clauset:2009} and Virkar and Clauset \cite{virkar2014} to our data to distinguish between power-law and lognormal distributions. Clauset \textit{et al.} \cite{Clauset:2009} approached the problem of estimating the exponent $\gamma$ in (\ref{pdf_powerlaw}) by testing different distributions. In their data, lognormals and power-laws were not clearly distinguishable either: for some tested data-sets, a lognormal distribution actually achieved a higher p-values (goodness of fit) than power-laws, but log ratio tests suggested that other tested distributions are closer to power-laws. By using the technique in \cite{Clauset:2009} to fit the distribution of $T_{friend}$, a power-law with exponent $\gamma=1.53$ was found with a reasonable $p$-value of $0.23$. The
distribution of $T_{follower}$ is most likely not a power-law because the $p$-value of $0.0$. Table
\ref{tab:EstimatedParametersUsingTheMethodOfClausetEtAl} lists the parameters for all used datasets using Clauset and Virkar's fitting technique \cite{virkar2014} with the according log-likelihood (Log-lh) and p values. A positive log-likelihood ratio indicates that the power-law is favored over the lognormal.

\begin{table}[th]
\hspace*{-2cm}
\begin{tiny}
\begin{tabular}
[c]{l|cll|ll|ll|r}
Distribution & 	 & power-law 	 &      & lognormal & 			 & Log-lh	& 		 & \# of data-points\\
(binned in)	 & Exponent ($\gamma$) & p & xmin & $\mu$ 				&$\sigma$& p 		& LR & \\\hline
&  &  &  &  &  & & & \\
adding friends $T_{friend}$ (seconds)& 1.53 & 0.23 & 59 &  0.54 & 2.26 & 0.0015 & 418.67 & 7,156,722\\
adding follower $T_{follower}$ (seconds) & 2.03 & 0.0 & 12 & 10.45 & 2.75 & 0 & -1327.5 & 6,734,405\\
adding follower $T_{follower}$ (hours) & 1.81 & 0.37 & 116 & 6.43 & 3.1 & 0 & -297.66 & 41,184\\
sending emails (seconds) & 2.33 & 0.0 & 86 & 9.53	& 3.24 & 0 & -373.7441 & 68,346,901\\
sending emails (minutes) & 2.03 & 0.21 & 1 & 6.9 & 3.32 & 0 & -391.13 & 62,031,701\\
receiving emails (seconds) & 2.01 & 0.0 & 1 & 10.31 & 2.73 & 0 & -958.67 & 69,528,701\\
receiving emails (minutes) & 2.21 & 0.05 & 182 & 6.23 & 3.54 & 0 & -136.31 & 15,365,001\\
commenting on Reddit.com (seconds) & 1.08 & 0.0 & 112 & 10.2 & 3.13 & 0 & -0.0024 & 180,156,539\\
commenting on Reddit.com (hours) & 1.91 & 0.01 & 64 & 0.88 & 2.41 & 0 & -20.73 & 50,044\\
&  &  &  &  &  & & & 
\end{tabular}
\end{tiny}
\caption{Estimated parameters using the method of Clauset \textit{et al.}
\cite{Clauset:2009}}%
\label{tab:EstimatedParametersUsingTheMethodOfClausetEtAl}%
\end{table}

\end{document}